\documentclass[preprint]{aastex6}
\begin{document}

\title{New Grids of Pure-Hydrogen White-Dwarf NLTE Model Atmospheres \newline
and the HST/STIS Flux Calibration}

\author{Ralph C. Bohlin\altaffilmark{1}, Ivan Hubeny\altaffilmark{2}, and
Thomas Rauch\altaffilmark{3}}
\altaffiltext{1}{Space Telescope Science Institute, 3700 San Martin Drive,
Baltimore,  MD 21218, USA}
\altaffiltext{2}{The University of Arizona, Steward Observatory, 933 North
Cherry Avenue, Tucson, AZ 85719, USA}
\altaffiltext{3}{Institute for Astronomy and Astrophysics, Kepler Center for
Astro and Particle Physics, Eberhard Karls University, Sand 1, 72076
T{\"u}bingen, Germany}

\begin{abstract} 

Non-local Thermodynamic Equilibrium (NLTE) calculations of hot white dwarf (WD)
model atmospheres are the cornerstone of modern flux calibrations for the Hubble
Space Telescope (HST) and for the CALSPEC database. These theoretical spectral
energy distributions (SEDs) provide the relative flux vs. wavelength, and only
the absolute flux level remains to be set by reconciling the measured absolute
flux of Vega in the visible with the Midcourse Space Experiment (MSX) values for
Sirius in the mid-IR. The most recent SEDs calculated by the \textsc{tlusty} and
\textsc{tmap} NLTE model atmosphere codes for the primary WDs G191-B2B, GD153,
and GD71 show improved agreement to 1\% from 1500~\AA\ to 30~\micron, in
comparison to the previous 1\% consistency only from 2000~\AA\ to 5~\micron.
These new NLTE models of hot WDs now provide consistent flux standards from the
FUV to the mid-IR.

\end{abstract}


\section{Introduction}			

Precise absolute flux calibration of astronomical spectra is often crucial for
understanding the nature of astronomical sources. White dwarf (WD) model
atmosphere calculations for three primary WD standards, viz. G191-B2B,  GD153,
and GD71, provide the basis for the Hubble Space Telescope (HST) and the
CALSPEC\footnote{http://www.stsci.edu/hst/instrumentation/reference-data-for-calibration-and-tools/astronomical-catalogs/calspec}
absolute flux scale \citep{bohlinetal14}. These models determine the shape of
the spectral energy distributions (SEDs), i.e., flux as a function of
wavelength, while \cite{bohlin14} set the absolute flux level by reconciling the
visible flux of Vega \citep{megessier95} at 5556~\AA\ (5557.5~\AA\ in vacuum)
with the MSX mid-IR fluxes \citep{price04} at 8--21\micron. \citet{bohlinetal14}
used Rauch \textsc{tmap}2012 models for the shape of the  SEDs, but the
uncertainties were defined by the discrepancy between the \textsc{tmap} and
\textsc{tlusty} pure hydrogen models for the three primary standards at the
$T_\mathrm{eff}$ and $\log g$ derived by \citet{gian2011}. \citet{bohlinetal14}
used their \textsc{tlusty204} version for these comparisons. New pure-hydrogen
WD model grids were computed by I. H. and by T. R. with greatly improved
agreement in comparison to the discussion in \citet{bohlinetal14}. In
conjunction with a metal-line blanketed model for G191-B2B, these new models
define revised SEDs for the three primary WD standards G191-B2B, GD153, and
GD71.

Section 2 presents the new WD grids and derives a lower error estimate from
the improved agreement of the two sets of NLTE calculations. Section 3 utilizes
the new models to revise the Space Telescope Imaging Spectrometer (STIS) flux
calibration, which is the primary basis for the CALSPEC database of flux
standards. Section 4 fits the new grids to STIS observations of a few WDs in
order to extend their flux estimates to 30~\micron.

\section{The New grids}			

Both I. H. and T. R. computed new grids of pure hydrogen WDs with the current
versions of the independent \textsc{tlusty207} and \textsc{tmap2019}
\citep{werner2003, rauch2003, werner2012, rauch13} NLTE software codes. The
latest public \textsc{tlusty} version is 205 \citep{hubeny2017a,hubeny1995,
hubeny2014}, but version 207 should be released by the time this article is
published. These grids are available\footnote{DOI 0.17909/t9-7myn-4y46} in the
Mikulski Archive for Space Telescopes (MAST). Both grids contain 132 models with
effective temperature ($T_\mathrm{eff}$) in the range 20,000-95,000~K and
surface gravity ($\log g$) between 7.0 and 9.5, with six steps of 0.5, where g
has units of $cm~s^{-2}$. The steps in $T_\mathrm{eff}$ are 2,000~K
between 20,000 and 40,000~K and 5,000~K between 40,000 and 95,000~K. Below
20,000~K, atmospheric convection becomes important and the computation of model
atmosphere SEDs is more complicated, e.g., \citet{gfusillo2020}.

\subsection{\textsc{tlusty} Grid}		

The model grid presented here is constructed by \textsc{tlusty207}, and the 
detailed spectra are synthesized by \textsc{synspec53}. The spectra cover the
vacuum wavelength range from 900~\AA\ to 32~\micron\ with a resolution R=5000.
The wavelength vector has 29712 sample points, which are the same for all
models. There is also a separate file with the theoretical continuum at a lower
resolution.  The individual entries represent the wavelength (\AA) and the
Eddington flux $H$ in units erg cm${}^{-2}$s${}^{-1}$\AA${}^{-1}$ at the stellar
surface. In order to account for limb darkening, the physical flux $F$ is the
integral over the outgoing 2$\pi$ steradians of the specific intensity; and $H$
is defined by $F_\lambda = 4\pi H_\lambda$. The naming is for example,
t400g800n, which means $T_{\rm eff} = 40,000$ K and $\log g = 8.00$, and n means
NLTE.

The model atmospheres are constructed using a 16-level model atom of hydrogen,
where the 16th level is the so-called ``merged level" (e.g.,
\citet{Hubeny2017b}, \S\,2.3) that represents all higher states lumped together.
For all levels, \citet{Hubeny1994} provide occupation probabilities. The
bound-free transitions from levels $n=1$ to $n=4$ are supplemented by
``pseudocontinua" that represent transitions to dissolved parts of the higher
levels. Their cross-sections are evaluated as described in \citet{Hubeny1994},
namely as an extrapolation of the regular photo-ionization cross-sections to
frequencies below corresponding thresholds, down to certain, essentially ad hoc,
cutoff frequencies. In the present models, these cutoffs are considered to be
free parameters and are set to $3.08 \times 10^{15}$ and $6.9 \times 10^{14}$
for the Lyman and Balmer pseudocontinua, respectively. The values of the cutoff
frequencies represent one of the few remaining uncertainties in the present DA
white dwarf model atmospheres. The line profiles for the first 20 lines of the
Lyman, Balmer, and Paschen series, and for the first 10 lines of the Brackett
series, are from \citet{tremblay2010}. The line profiles of the higher members
of these series and of the higher series are taken from Kurucz's ATLAS
\citep{Kurucz1970}. For Lyman $\alpha$, $\beta$, and $\gamma$, we have actually
used only a half of the Tremblay-Bergeron value that corresponds to broadening
by electrons; the broadening by protons was replaced by the broadening theory
that includes quasi-molecular satellites of these lines, after
\citet{allard1994}, although there is some evidence that the \citet{allard1994}
formulation over-estimates their strengths.

\begin{figure}			
\centering 
\includegraphics*[width=.9\textwidth,trim=0 35 0 0]{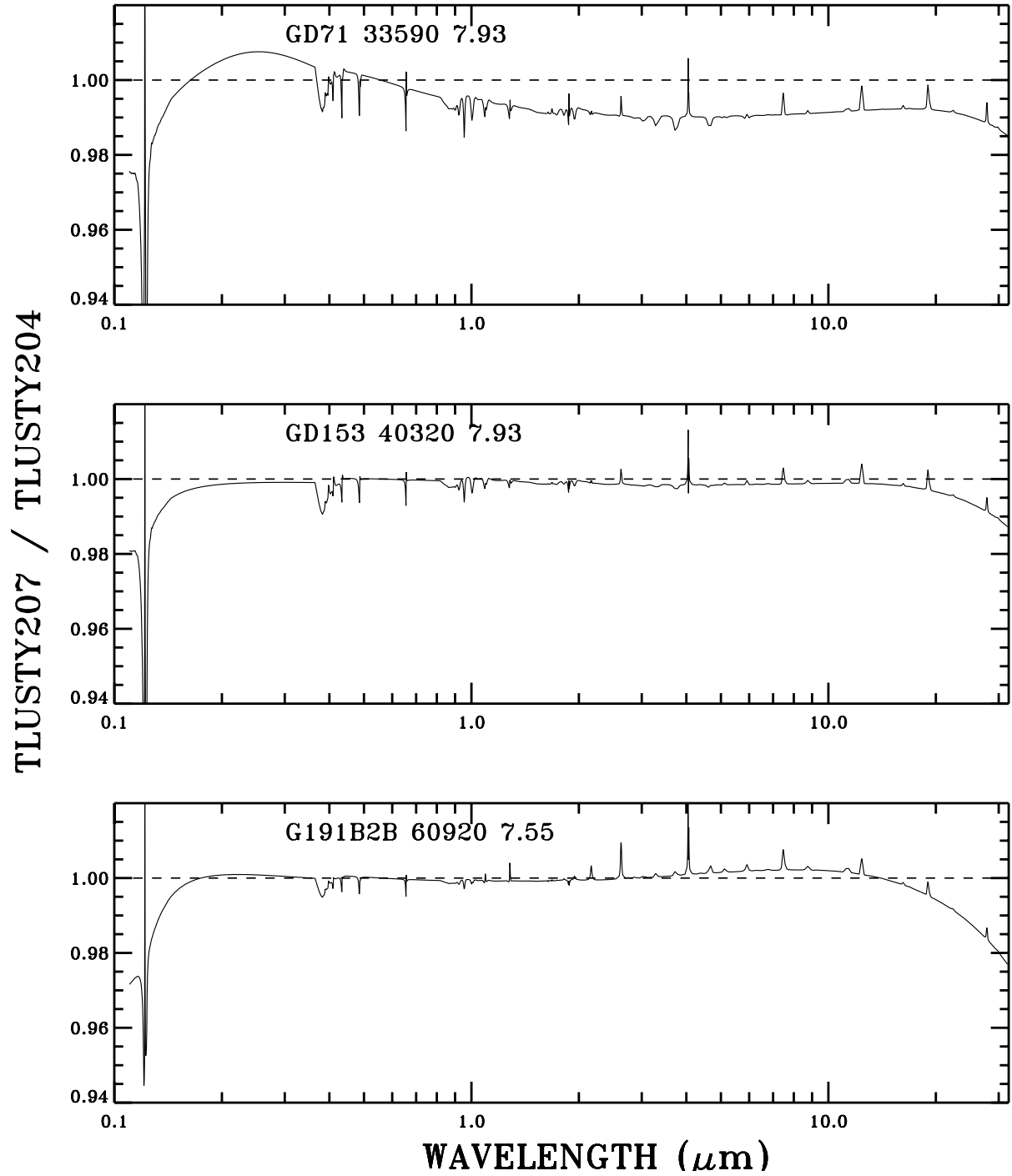}
\caption{\baselineskip=12pt
Change in the pure hydrogen model SEDs for the three primary standards between
the \textsc{tlusty204} calculations used in \citet{bohlinetal14} and new
\textsc{tlusty207} models. Both the numerator and denominator SEDs are at the
\citet{gian2011} $T_\mathrm{eff}$ and $\log g$ used by \citet{bohlinetal14} and
written on each panel of the Figure. \label{fig1}} \end{figure}

Figure~\ref{fig1} compares the \textsc{tlusty204} models from
\citep{bohlinetal14} to the new \textsc{tlusty207} computations for pure
hydrogen at the same \citep{gian2011} $T_\mathrm{eff}$ and $\log g$ for the
three primary standard WDs. The changes are mostly less than 1\% but range to
3\% at 1100~\AA\ and to 2\% at 30~\micron\ for the hottest model. The
differences are due to a newer version of the Tremblay-Bergeron line broadening
tables, kindly supplied by P.E. Tremblay \citep[][extended tables of 2015,
priv\@. comm\@.]{tremblay2009}. Version 204 used an older version of
the tables, where only the first 10 members of the Lyman and Balmer series were
treated. Other lines profiles use an approximation described in Appendix B of
\citet{Hubeny1994}. There are also some slight changes in a treatment of the
Lyman and Balmer pseudocontinua.

\textsc{tlusty207} is also used to compute a NLTE metal-line blanketed model
atmosphere for G191-B2B with metal abundances from \citet{rauch13}. All H, He,
C, N, O, Al, Si, P, S, Fe, and Ni species are treated explicitly in NLTE, while
Ti, Mn, Cr, Cu, and Zn are treated in LTE. The effective temperature and surface
gravity are $T_{\rm eff} = 59,000$ K, and $\log g = 7.60$, according to
\citet{bohlinetal14}. The number of explicit levels and superlevels for the
individual ionization stages of the explicit atoms and the corresponding atomic
data are similar to those considered in the model atmosphere grids OSTAR2002
\citep{Lanz03} and BSTAR2006 \citep{Lanz07}. The hydrogen atom and hydrogen
lines are treated in the same manner as in the main grid of pure-H  DA white
dwarfs described above. 

Using this \textsc{tlusty} model atmosphere for G191-B2B, \textsc{synspec53}
produced two synthetic spectra: a detailed spectrum with all metal lines between
900 and 2000~\AA\ at a nominal resolution $R=200,000$; and for the longer
wavelengths, a ``standard" synthetic spectrum at $R=5000$ with only hydrogen and
helium lines. In order to make a complete spectrum for use as a standard
star SED, these two pieces are concatenated for full wavelength coverage.  The
model computation for both pieces uses the same self-consistent structure that
takes into account the effect of metal lines and NLTE effects in all considered
species. The only difference for the long wavelength segment is the omission of
metal lines, which are very weak and are not used for a spectroscopic analysis.
A high-resolution synthetic spectrum in the whole frequency range would be an
unnecessary overkill. As evidence of the validity of the composite, the
continuum above and below 2000~\AA\ match without adjustment, making a seamless
join at 2000~\AA. This composite \textsc{tlusty207} model is in CALSPEC as
\textit{g191b2b\_mod\_011.fits}.

\subsection{\textsc{tmap} Grid}			

The software version is \textsc{tmap2019}, and the spectra cover the vacuum
wavelength range from 900~\AA\ to 30~\micron\ with a variable resolution that is
higher in the lines and ranges upward of R=50,000. The wavelength vector has
144,796 sample points, which are the same for all models. There are four columns
to each model, i.e., wavelength in \AA, the  \textit{astrophysical surface
flux}, i.e., $4H_\lambda$ in erg~cm$^{-2}$~s$^{-1}$~cm$^{-1}$, the ratio of the
flux to the theoretical continuum level, and the continuum in the same units as
the flux. These units make the \textsc{tmap2019} units $4\times10^{8}$ larger
than the \textsc{tlusty} units, $H_\lambda$, after accounting for the conversion
of cm$^{-1}$ to \AA${}^{-1}$.  To get the physical flux, $F_\lambda$, in
the common erg~cm$^{-2}$~s$^{-1}~$\AA$^{-1}$ units multiply the
\textsc{tmap2019} units by $\pi\times10^{8}$, instead of the $4\pi$ required for
the \textsc{tlusty207} conversion. These conversion factors are necessary for
computing stellar angular diameters from dereddened CALSPEC fluxes and the
measured parallax distance.

The naming is for example, 0040000\_8.00, which means $T_\mathrm{eff}$=40,000~K
and $\log g$=8.00. These \textsc{tmap} models are available via the Theoretical
Stellar Spectra Access (TheoSSA\footnote{\url{http://dc.g-vo.org/theossa}})
service, which has been created in the framework of the T\"ubingen project of
the German Astrophysical Virtual Observatory
(GAVO\footnote{\url{http://dc.g-vo.org}}) and provides easy access to TMAP SEDs.
This archive includes several standard stars, e.g., G191-B2B, GD\,71, and
GD\,153.

\begin{figure}			
\centering 
\includegraphics*[width=.99\textwidth,trim=0 35 0 0]{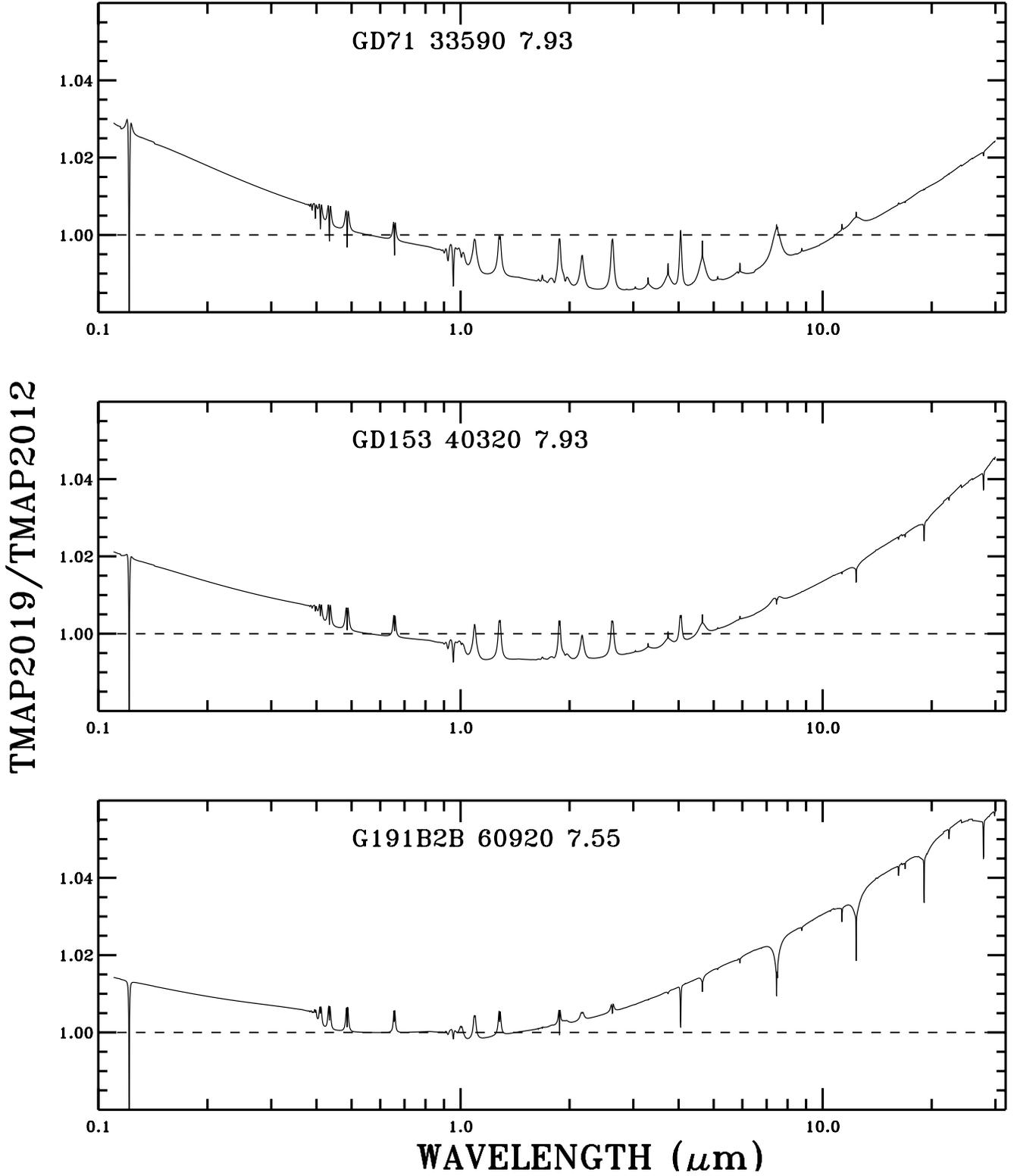}
\caption{\baselineskip=12pt
Change in the pure hydrogen model SEDs for the three primary standards between the \textsc{tmap}2012 calculations used in \citet{bohlinetal14} and new \textsc{tmap2019} models. Both the numerator and denominator SEDs are at the \citet{gian2011}
$T_\mathrm{eff}$ and $\log g$ used by \citet{bohlinetal14} and written on each
panel.
\label{fig2}} \end{figure}

Figure~\ref{fig2} compares the \textsc{tmap}2012 models from
\citep{bohlinetal14} to the new \textsc{tmap2019} computations for pure hydrogen
at the same \citep{gian2011} $T_\mathrm{eff}$ and $\log g$. The changes range
from 3\% at 1100~\AA\ for the coolest model to 6\% at 30~\micron\ for the
hottest model. The \textsc{tmap} code was modified between 2012 and the current
2019 version to match \textsc{tlusty207} in the long cutoff frequency used for
the  H I Lyman bound-free opacity of the H I ground-state absorption
\citep{rauch2008}. \textsc{tmap2019} uses a 15 level H I model ion that also
matches \textsc{tlusty207}.

\begin{figure}			
\centering 
\includegraphics*[width=.99\textwidth,trim=0 35 0 0]{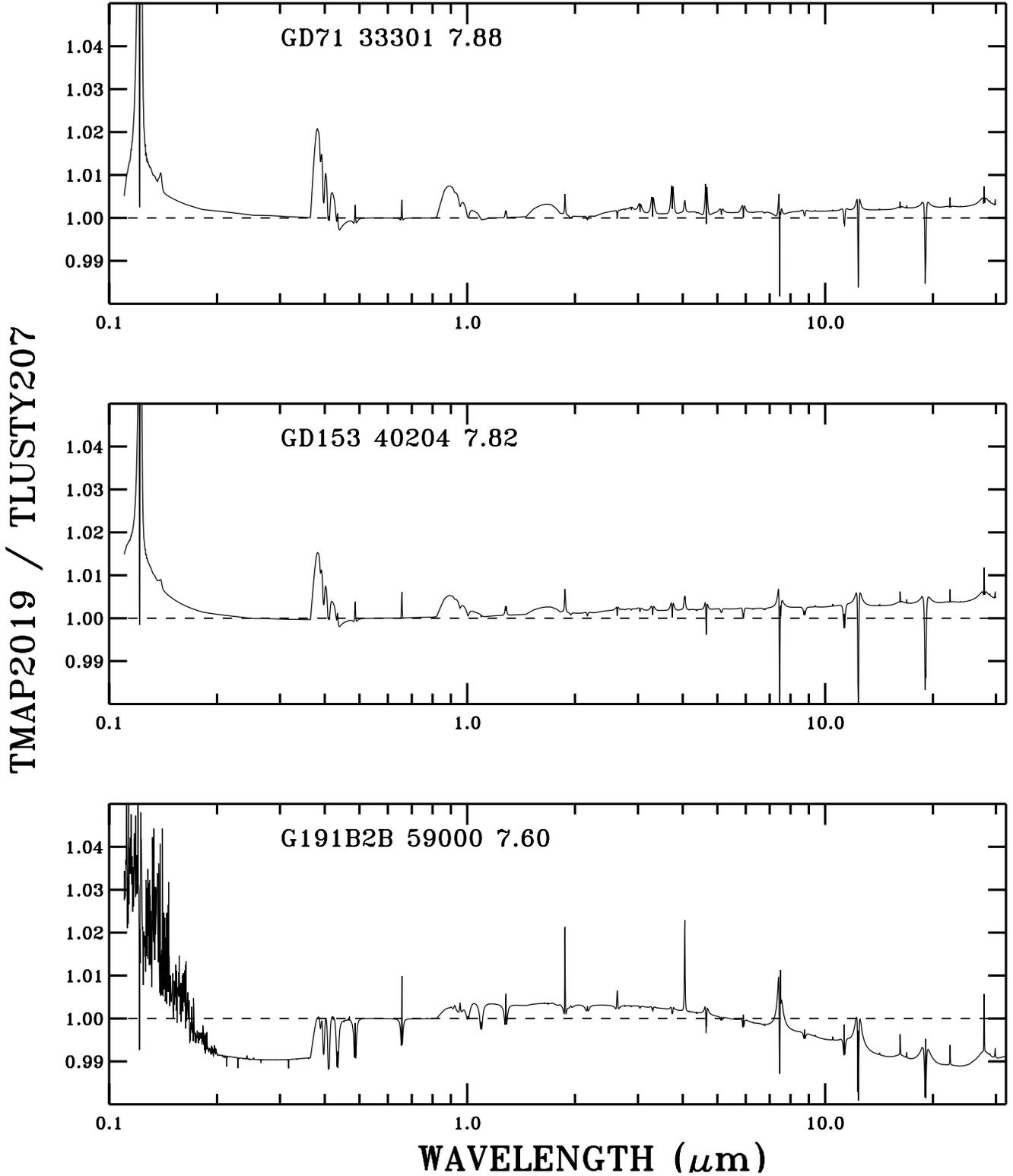}
\caption{\baselineskip=12pt
Comparison of the new \textsc{tmap2019} and \textsc{tlusty207} NLTE models for the three
primary standards. Both the numerator and denominator SEDs are at the
average $T_\mathrm{eff}$ and $\log g$ of Table~\ref{table:tg} for the two pure hydrogen WDs,
while the bottom panel compares line blanketed models for G191-B2B at the same
$T_\mathrm{eff}$=59,000~K and $\log g=7.60$. The model ratios are normalized to
unity at 5556~\AA\ (5557.5~\AA\ in vacuum). The uncertainty relative to 5557.5~\AA\
in the HST/CALSPEC flux scale is partially defined by the rms of these
differences. Because the \textsc{tmap2019} model is unchanged from the 2013 version used by \citet{bohlinetal14}, the bottom panel represents the change in the reference SED for G191-B2B, i.e., the change in the CALSPEC reference file from g191b2b\_mod\_010.fits to g191b2b\_mod\_011.fits.
\label{fig3}} \end{figure}

\begin{deluxetable}{cccccccc}		
\tablewidth{0pt}
\tablecolumns{8}
\tablecaption{\label{table:tg} The Primary WD Stars}
\tablehead{
\colhead{Star} &\colhead{V} &\colhead{Sp. T.}  &\colhead{$E(B-V)$} &\colhead{$T_\mathrm{eff} (K) $} &\colhead{$\log g$}
&\colhead{Unc. $T_\mathrm{eff} (K) $} &\colhead{Unc. $\log g$}}
\startdata
G191-B2B &11.781\tablenotemark{a} &DA.8  &0.0005 &59000 &7.60  &2000 &0.05\\
GD153   &13.349\tablenotemark{b} &DA1.2 &0.0002 &40204 &7.82  &585 &0.05\\
GD71    &13.032\tablenotemark{c} &DA1.5 &0.0001 &33301 &7.88  &342 &0.04\\
\enddata
\tablenotetext{a}{\citet{landoltuom07}}
\tablenotetext{b}{\citet{bohlinarlo15}}
\tablenotetext{c}{\citet{Landolt92}}
\tablecomments{Columns 7 and 8 are the uncertainties.}
\end{deluxetable}

\subsection{Comparisons and Uncertainties}		

Since the \citet{gian2011} analysis, \citet{narayan2019} redetermined 
$T_\mathrm{eff}$ and $\log g$ for the pure hydrogen GD153 and GD71 using new
observations of the Balmer line profiles. Because the independent
\citet{gian2011} and \citet{narayan2019} analyses have comparable uncertainties,
an average of these two sets of determinations should have a reduction of the
error bars by $\approx\sqrt{2}$. To establish conservative uncertainties on the
averages, the largest of the Gianninas and Narayan upper and lower limits is
reduced by $\sqrt{2}$ for our one sigma uncertainty estimates. For example,
these uncertainties for the $T_\mathrm{eff}$ of GD153 are 626~K for Gianninas
and (+827, -497)~K from Narayan, so our average $T_\mathrm{eff}$ and
uncertainty in Table~\ref{table:tg} are 40204~K and $\pm827/\sqrt{2}$=585~K.
Both independent $T_\mathrm{eff}$ determinations for GD153 and GD71 are within
1$\sigma$ of our average. Table~\ref{table:tg} lists these values along with the
$T_\mathrm{eff}$=59,000~K and $\log g=7.60$ for the blanketed G191-B2B model
\citep{bohlinetal14} with its 2000~K and 0.05 $\log g$ uncertainties of
\citet{rauch13}.

Figure~\ref{fig3} compares the \textsc{tmap2019} to the \textsc{tlusty207}
models for the Table~\ref{table:tg} parameters. Special models for GD153 and
GD71 with the Table~\ref{table:tg} parameters can avoid small errors of $<$0.1\%
from interpolation in the grids. The \textsc{tlusty} and \textsc{tmap} NLTE
models agree to 1\% over most of the range longward of 1500~\AA\ but differ by
1--3\% at the 1150~\AA\ short wavelength cutoff of the STIS sensitivity.
Previously, in \citet{bohlinetal14}, there was similar agreement to 1\% only at
0.2--5~\micron.  When the \textsc{tlusty207} and \textsc{tmap2019} model
calculations have the same basic parameters, including the same metal
abundances, the agreement is good, regardless of a different treatment of
metals, number of NLTE levels, organization of Fe and Ni levels into
superlevels, and detailed atomic parameters for metals.

Also included in Table~\ref{table:tg} are the selective extinction E(B-V) values
computed from the interstellar neutral hydrogen column densities of
$N(HI)=2.2\times10^{18}$ \citep{rauch13}, $0.98\times10^{18}$, and
$0.63\times10^{18}$ \citep{dupuis1995} for G191-B2B, GD153, and GD71,
respectively. The Galactic average of $N(HI)/E(B-V)=4.8\times10^{21}$ of
\citet{bohlin1978} determines the extinction E(B-V) to within a factor of two
for the typical uncertainty. Using the LMC reddening curve \citep{koornneef1981}
for the small grains in the local interstellar medium, the maximum reddening
from the dust is 0.65\% at 1150~\AA\ or 0.5\% at the 1674~\AA\ short wavelength
limit of STIS coverage for G191-B2B.

\begin{figure}
\centering 
\includegraphics*[width=.99\textwidth,trim=0 15 0 0]{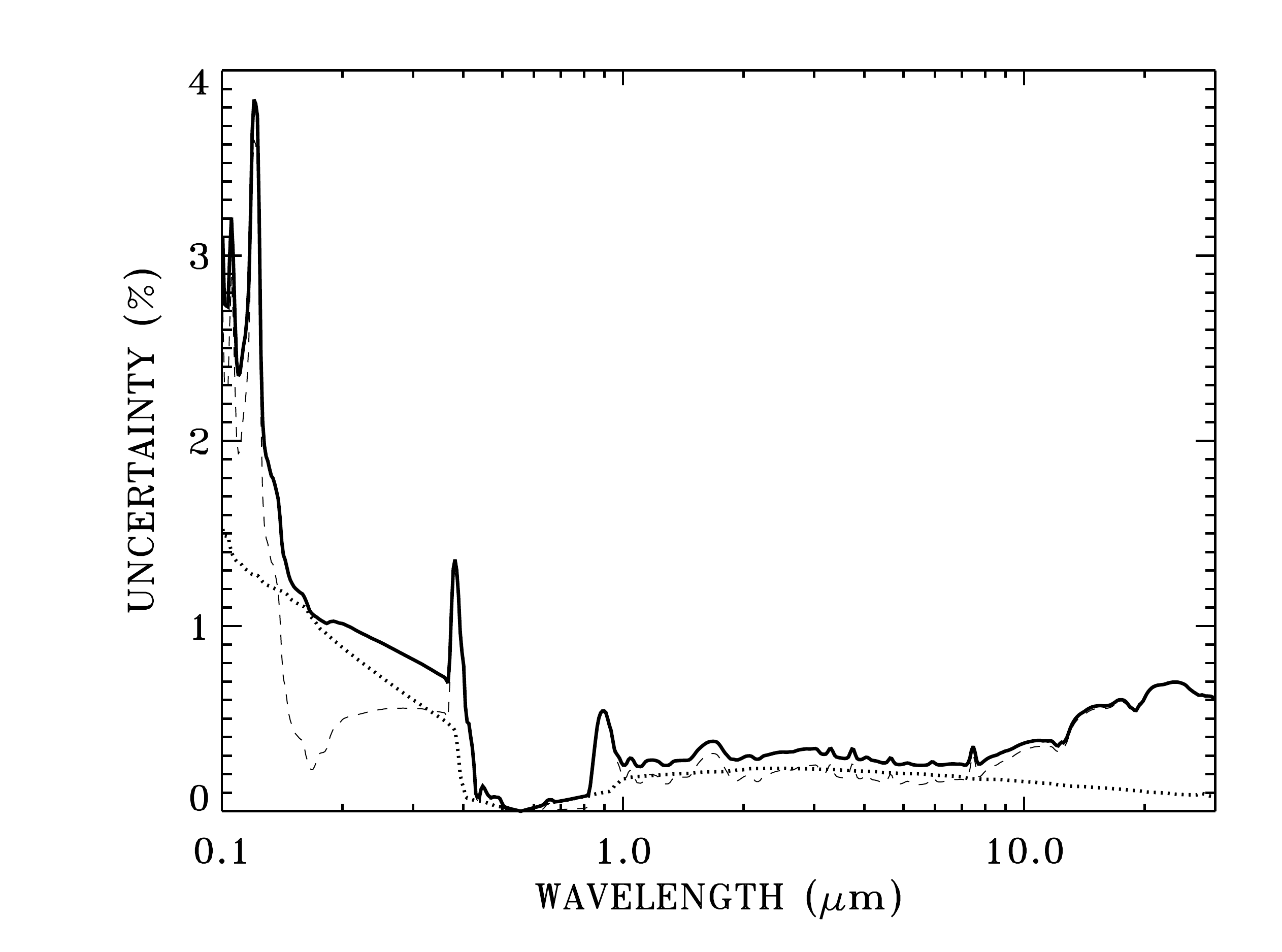}
\caption{\baselineskip=12pt
Formal uncertainty relative to 5557.5~\AA\ in the HST/CALSPEC WD flux scale.
\textit{Dotted line:} uncertainty attributed to the determination of 
$T_\mathrm{eff}$ and $\log g$ from Table~\ref{table:tg} for the three primary
standard WDs combined in quadrature.
\textit{Dashed line:} rms uncertainty attributed to the differences in the \textsc{tmap} and
\textsc{tlusty} models from Figure~\ref{fig3}. \textit{Heavy solid line:} total uncertainty from the combination in quadrature of the above two independent uncertainties.
\label{uncert}} \end{figure}

Following the procedure in the review paper \citep{bohlinetal14}, the rms of the
differences in Figure~\ref{fig3} define the uncertainty in the HST/CALSPEC flux
scale when the formal uncertainties in each model are also included. For the
model uncertainties of G191-B2B, a 61,000~K $\log g$=7.65 metal line-blanketed
model is compared to the baseline 59,000~K $\log g$=7.60 from
Table~\ref{table:tg}. Figure~\ref{uncert} shows these two separate contributions
and their combination in quadrature to get the total uncertainty as a function
of wavelength. The \textsc{tmap} and \textsc{tlusty} models agree so well that
the uncertainties in the $T_\mathrm{eff}$, $\log g$ determinations dominate from
1300--3700~\AA. The peaks in the uncertainty correspond to the remaining
modeling uncertainties in hydrogen opacities at Ly$\beta$ and Ly$\alpha$ and at
the confluences of the Balmer and Paschen lines. The formal uncertainty in the
IR remains below 1\% in contrast to the similar figure from
\citet{bohlinetal14}, where there is a total error bar of 4\% at 30~\micron. To
find the total absolute uncertainty in HST/CALSPEC fluxes, the 0.5\% uncertainty
of the absolute 5557.5~\AA\ flux level (see the next Section) must be added in
quadrature to the Figure~\ref{uncert} uncertainty relative to 5557.5~\AA. An
updated covariance matrix \citep{bohlinetal14} for the uncertainties is in
CALSPEC.

\newpage

\section{Recalibration}			

\begin{figure}			
\centering 
\includegraphics*[width=.99\textwidth,trim=0 65 0 85]{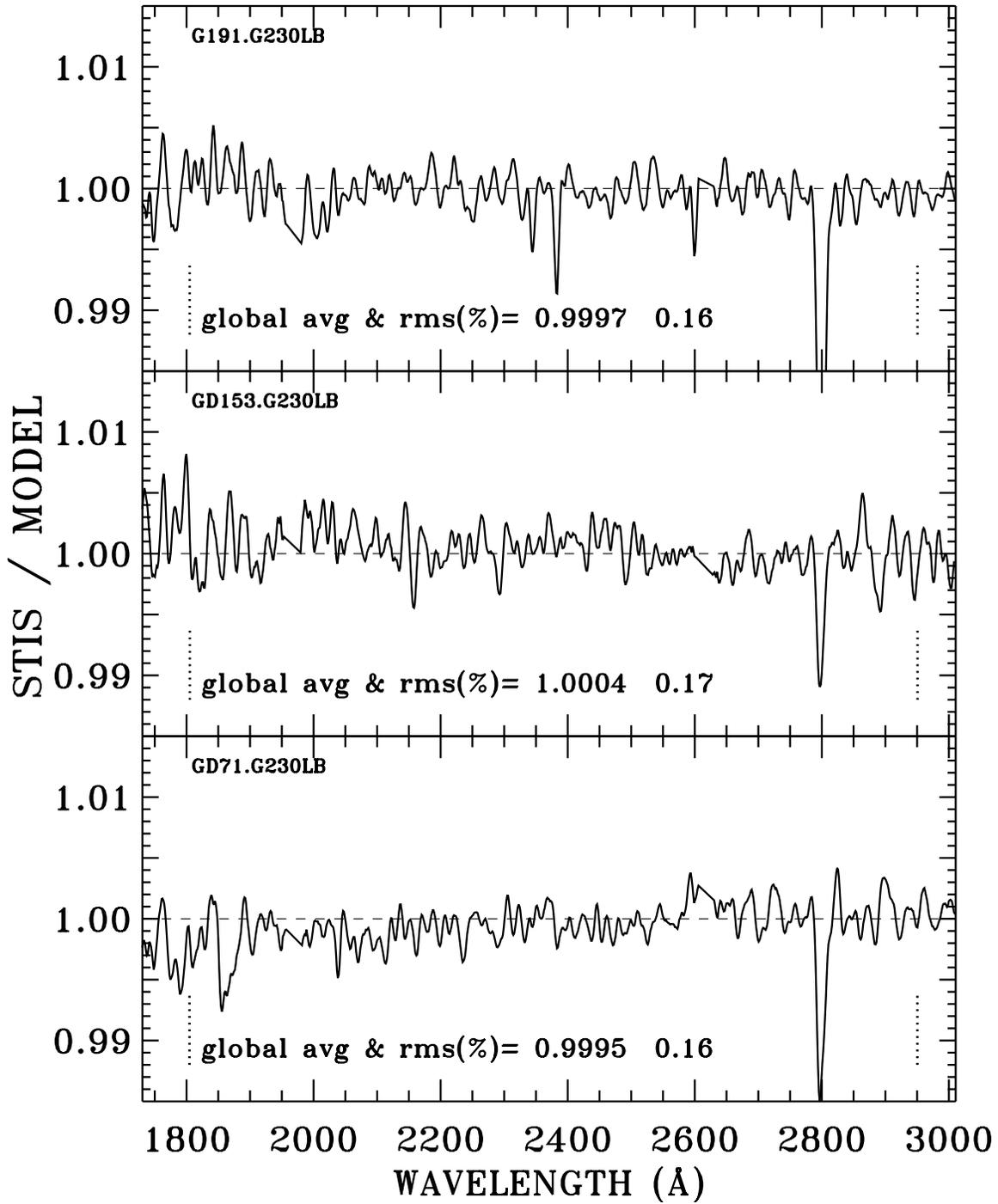}
\caption{\baselineskip=12pt
\textsc{tlusty} residuals, i.e., the ratios of the STIS flux from a calibration
using \textsc{tlusty} models divided by the \textsc{tlusty} model flux distributions. The average ratio and its rms scatter between the vertical dotted lines is written on each of the three panels. The feature at 2800~\AA\ is the 
MgII interstellar line, which is masked and does not affect the flux calibration. Notice the tyical broadband consistency of the flux calibration to 0.1--0.2\% among the three stars.
\label{residhub}} \end{figure}

\begin{figure}			
\centering 
\includegraphics*[width=.99\textwidth,trim=0 65 0 85]{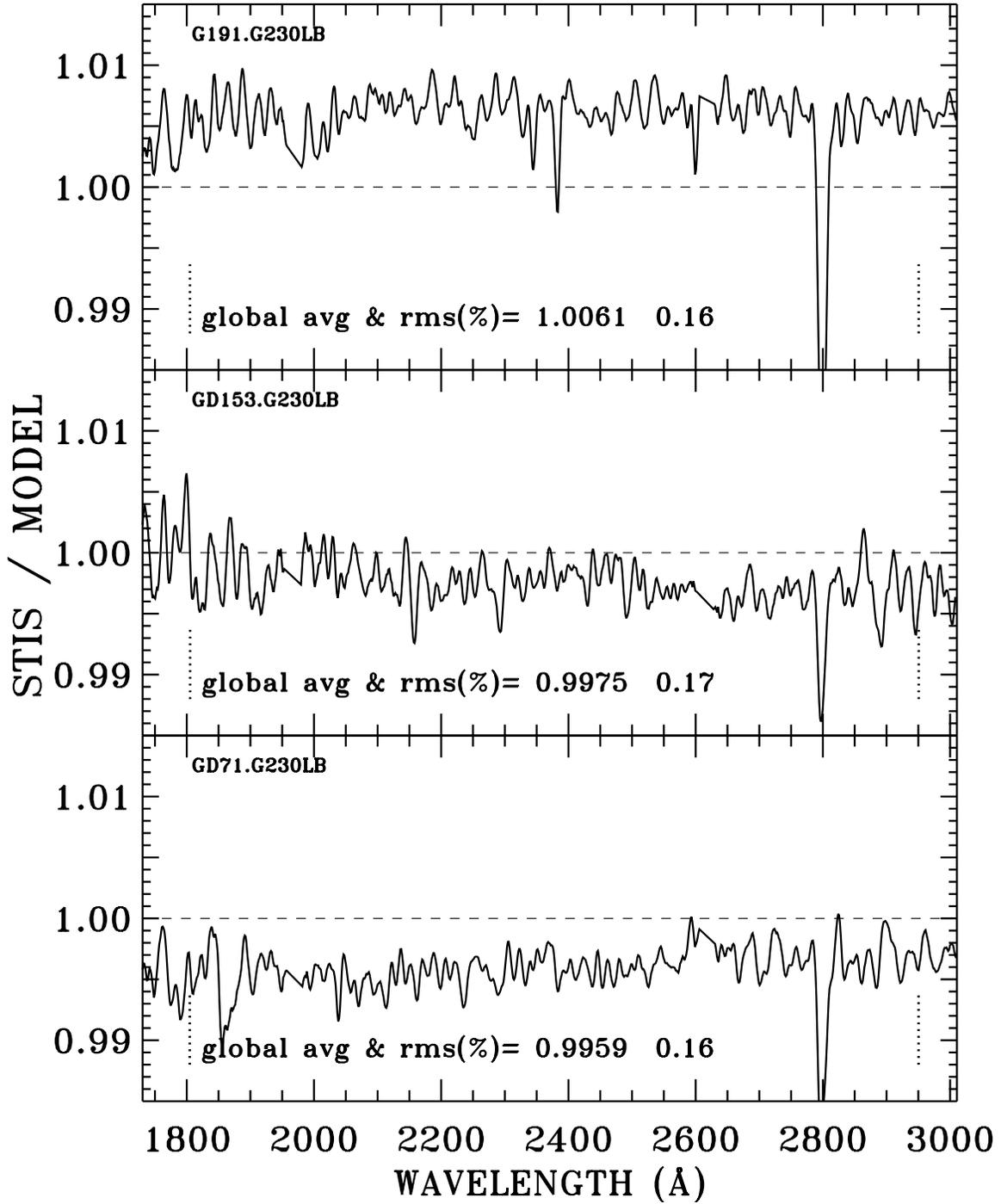}
\caption{\baselineskip=12pt
\textsc{tmap} residuals, i.e., the ratios of the STIS flux from a calibration
using \textsc{tmap} models divided by the \textsc{tmap} model flux
distributions. Notice the $\approx$1\% offset of the G191-B2B ratio in comparison
with the GD153 and GD71 ratios, which reflects the inconsistency between the
\textsc{tmap} model SED for G191-B2B and its STIS signal in comparison to same
ratio for the other two stars. \label{residrau}} \end{figure}

The general principles of instrumental flux calibration are documented by 
\citet{bohlinetal14}, while \citet{Bohlin2019} explain the calibration procedure
and its precision for the STIS low-dispersion spectral modes. Figure~\ref{fig3}
shows a systematic difference of 1\% from 2000-3900~\AA\ between the two
G191-B2B models, while the models for the other two stars differ by much less.
Which G191-B2B model is preferred in this wavelength range is determined by 
which one produces the smaller residuals in comparison to the STIS observations.
Figures~\ref{residhub}--\ref{residrau} show these residuals for the G230LB
spectral region for flux calibrations based on the respective \textsc{tlusty}
and \textsc{tmap} sets of models. While broadband systematic differences among
the three stars are limited to $\approx$0.2\% for the \textsc{tlusty} set, the 
\textsc{tmap} model for G191-B2B produces a calibration that is discrepant by
$\approx$1\% from the other two stars. 

\begin{figure}			
\centering 
\includegraphics*[width=.99\textwidth,trim=0 35 0 0]{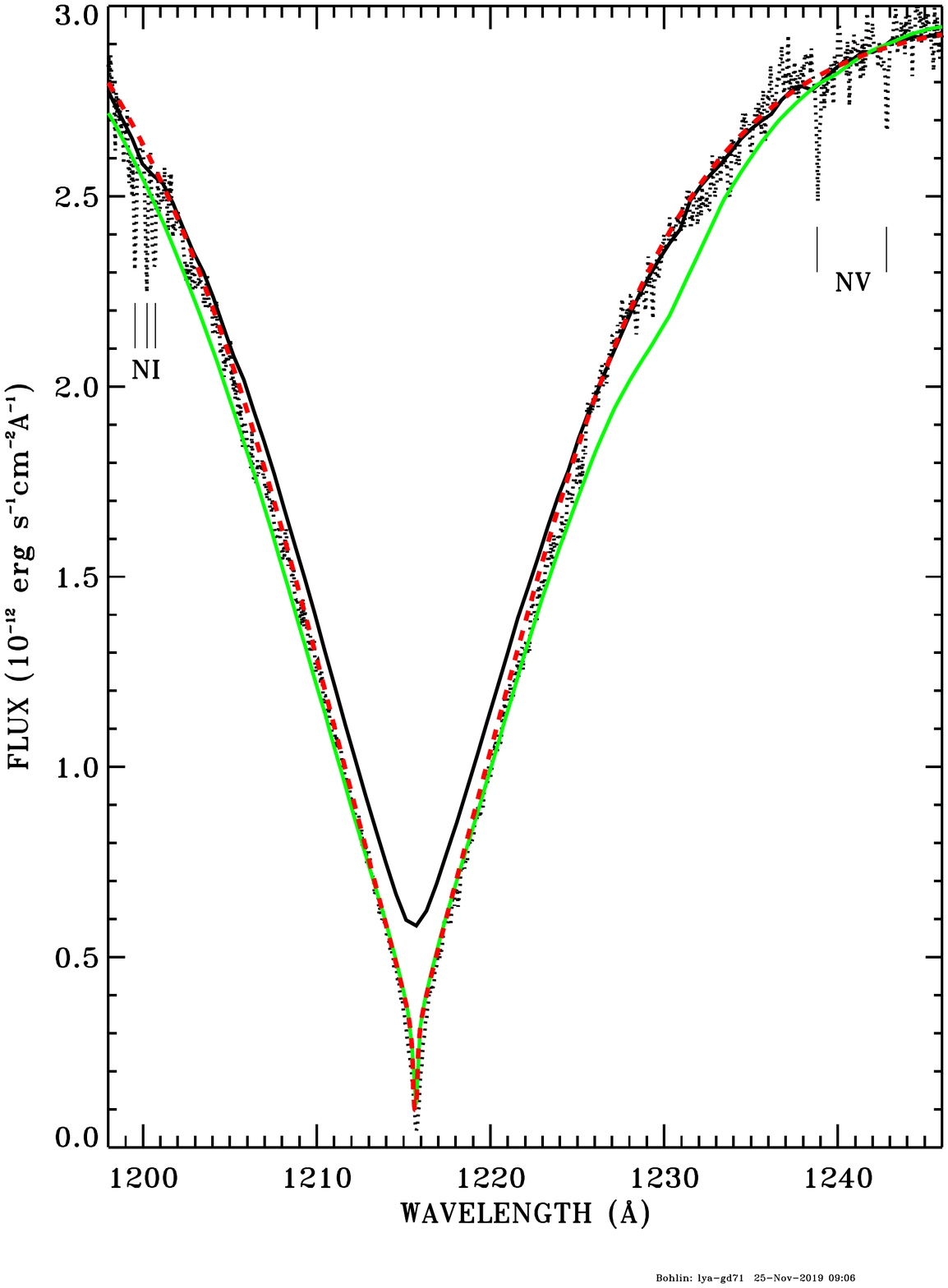}
\caption{\baselineskip=12pt
Four Ly$\alpha$ profiles for GD71: \textit{Heavy black line}: STIS G140L low
dispersion at R$\approx$500, \textit{dots}: STIS G140M (o4sp010b0) at R$\approx$5000,
\textit{green}: \textsc{tlusty} model, \textit{red dash}: \textsc{tmap} model. The data have been corrected to rest wavelengths using the radial velocity of
23.4~km s$^{-1}$. In addition, the G140M flux vector is adjusted for a $\approx$5\%
calibration error longward of 1225~\AA\ by normalizing to the G140L fluxes. The 
\textsc{tlusty} and \textsc{tmap} models are both normalized to the data in the
long-wavelength wing of the profiles. The positions of three NI
lines and two NV lines in the G140M data are marked with vertical
lines. \label{lya}} \end{figure}

Another difference between the two NLTE results is the inclusion of
quasi-molecular hydrogen lines in \textsc{tlusty} but not yet in the
\textsc{tmap} code. These quasi-molecular features are most pronounced in the
coolest star and their main impact lies below the 1150~\AA\ STIS cutoff;
however, a subtle flux depression appears at $\approx$1230~\AA\, as illustrated
for GD71 in Figure~\ref{lya}. Although there is some evidence for a much weaker
G140L feature, which is slightly below the red \textsc{tmap} profile, the green
\textsc{tlusty} feature is stronger than supported by the observations, which
suggests that the \textsc{tmap} models are preferred at Ly$\alpha$. The G140M
absorption line regions are masked for the G140L flux calibration. In
Figure~\ref{lya}, the G140L line-core level is raised by $\approx$10\% of the
continuum level by stray light from the broad wings of the wide line-profile
with the 52$\times$2 slit; however, the narrow-slit (0.2$\times$0.2), higher-resolution G140M
line-core agrees remarkably well with the models. The region of contamination
in the low-dispersion G140L line core is too wide for a successful spline fit
between the unmasked wings; and the sensitivity must be estimated by linear
interpolation between 1200 and 1230~\AA\ for the \textsc{tmap} models. The
extraneous \textsc{tlusty} quasi-molecular feature would require even more
interpolation over 1200--1240\AA.

In summary, the final set of models for the three primary standard stars
includes the pure hydrogen \textsc{tmap} models for GD153 and GD71 that have no
quasi-molecular hydrogen features in agreement with the observations.  For
G191-B2B, the \textsc{tlusty} model produces a more consistent flux calibration
than the \textsc{tmap} model in the 2000-3900~\AA\ range. This set of one
\textsc{tlusty} and two \textsc{tmap} models forms the primary set of standard
candles that is used for the flux calibration, which is defined by the average
of the sensitivities from the comparably robust sets of STIS observations for
each of the three primary standards. The sensitivities are the ratios of the
observed signals to the reference SEDs which are defined by  normalizing each
model to its stellar flux F in the 25~\AA\ band centered at 5557.5~\AA\ (vac),
where F = F(Vega)~R. The ratio R is the relative brightness of each primary to
Vega, as determined by  the average STIS signal for each of the WD primaries
divided   by the STIS Vega signal in the same 25~\AA\ band. F(Vega) =
$3.47\times10^{-9}$~erg cm$^{-2}$ s$^{-1}$ \AA$^{-1}$ at 5557.5~\AA\ (see the
next Section). There is no dependence on the photometry for Vega or any of the
primary standards, in contrast to the original normalization of the models using
V-band photometry \citep{bohlin2000}.

The STIS flux calibration observations  all utilize the wide photometric 52x2
arcsec slit, but the point spread function (PSF) has extensive wings that bring
contaminating continuum flux into the absorption line cores, especially for MAMA
observations. Thus, the line cores are masked where the wavelength regions in
Table~\ref{table:masks} must be ignored in the derivation of the sensitivity
vector, which is defined by a spline fit with a node spacing of 5~\AA\ at the
short wavelengths and 200~\AA\ at the longest wavelengths.

\newpage

\begin{deluxetable}{cc}			
\tablewidth{0pt}
\tablecolumns{2}
\tablecaption{\label{table:masks} Masked Wavelength Regions}
\tablehead{\colhead{ID} &\colhead{Range~(\AA)}}
\startdata
NI\tablenotemark{a}   &1198--1202.2\\
Ly$\alpha$            &1209--1222\\
NV\tablenotemark{a}   &1238--1239.6\\
OI\tablenotemark{a}   &1300--1304.4\\
CII\tablenotemark{a}  &1332--1338\\
CIV\tablenotemark{a}  &1547--1552\\
MgII\tablenotemark{a} &2790--2810\\
H$\zeta$              &3885--3905\\
H$\epsilon$           &3955--3990\\
H$\delta$             &4070--4140\\
H$\gamma$             &4332--4352\\
H$\beta$              &4857--4870\\
H$\alpha$             &6552--6577\\
Pa$\epsilon$           &9500--9600\\
Pa$\delta$             &10000--10100\\
\enddata
\tablenotetext{a}{Interstellar/Circumstellar}
\end{deluxetable}

\begin{figure}			
\centering 
\includegraphics*[width=.99\textwidth,trim=0 35 0 55]{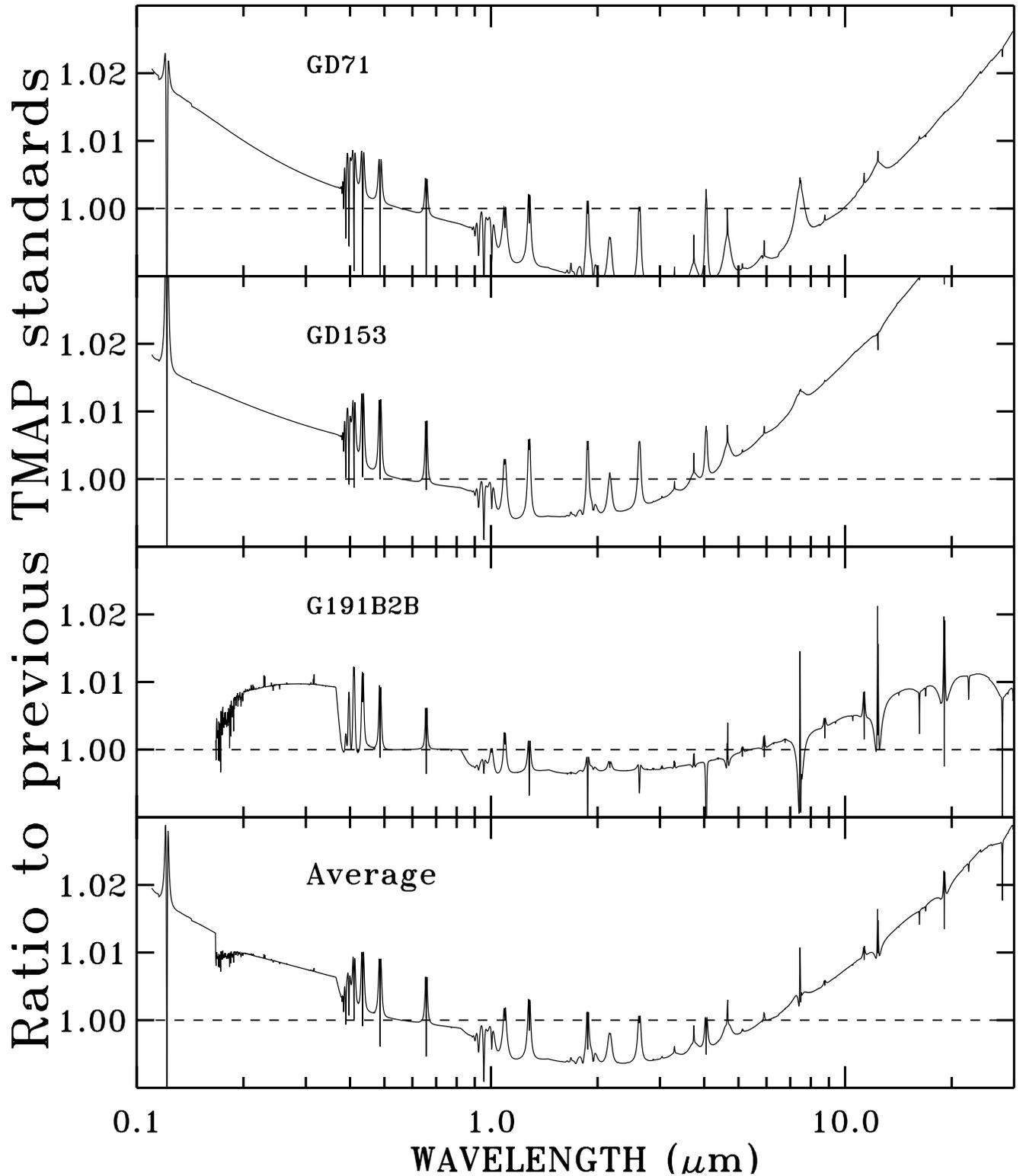}
\caption{\baselineskip=12pt
Change in the SED for each of the three primary flux standards and the average
change due to changes in the reference models. The average corresponds to the
change in flux for any calibration that is based on an equal weighting of the
three standard candles above the short wavelength G230LB cutoff at 1670~\AA.
G191-B2B is too bright for the MAMA G140L and G230L modes, so the flux
calibration of those two modes depends only on GD153 and GD71. The average
exceeds 1\% only longward of 10~\micron\ and shortward of 1670~\AA. Because the
absorption lines are masked in the flux calibration process, the illustrated
sharp spikes at line centers have no effect on the results, i.e., the measured
STIS flux distibutions.  \label{flxchang}} \end{figure}

Figure~\ref{flxchang} illustrates the change in the HST/CALSPEC flux for each of
the three primary WD standards. For a flux calibration that is based on equal
weights for the three primary stars, the change in the flux calibration would be
just the average of the three top panels. However, G191-B2B is too bright for the
STIS pulse counting G140L and G230L MAMA detectors; and the STIS fluxes for
these two modes are determined entirely by the pure hydrogen models for GD153
and GD71. G191-B2B can be used for all three CCD modes, which cover the
1670--10200~\AA\ range. Thus, the average change in the bottom panel represents
the average for all three stars above 1670~\AA\ and for the two fainter stars
below 1670~\AA. These average changes that are due to changes in the reference WD models are $<$1\% from 1670~\AA\ to 10~\micron. 

\begin{figure}			
\centering 
\includegraphics*[width=.99\textwidth,trim=0 30 0 55]{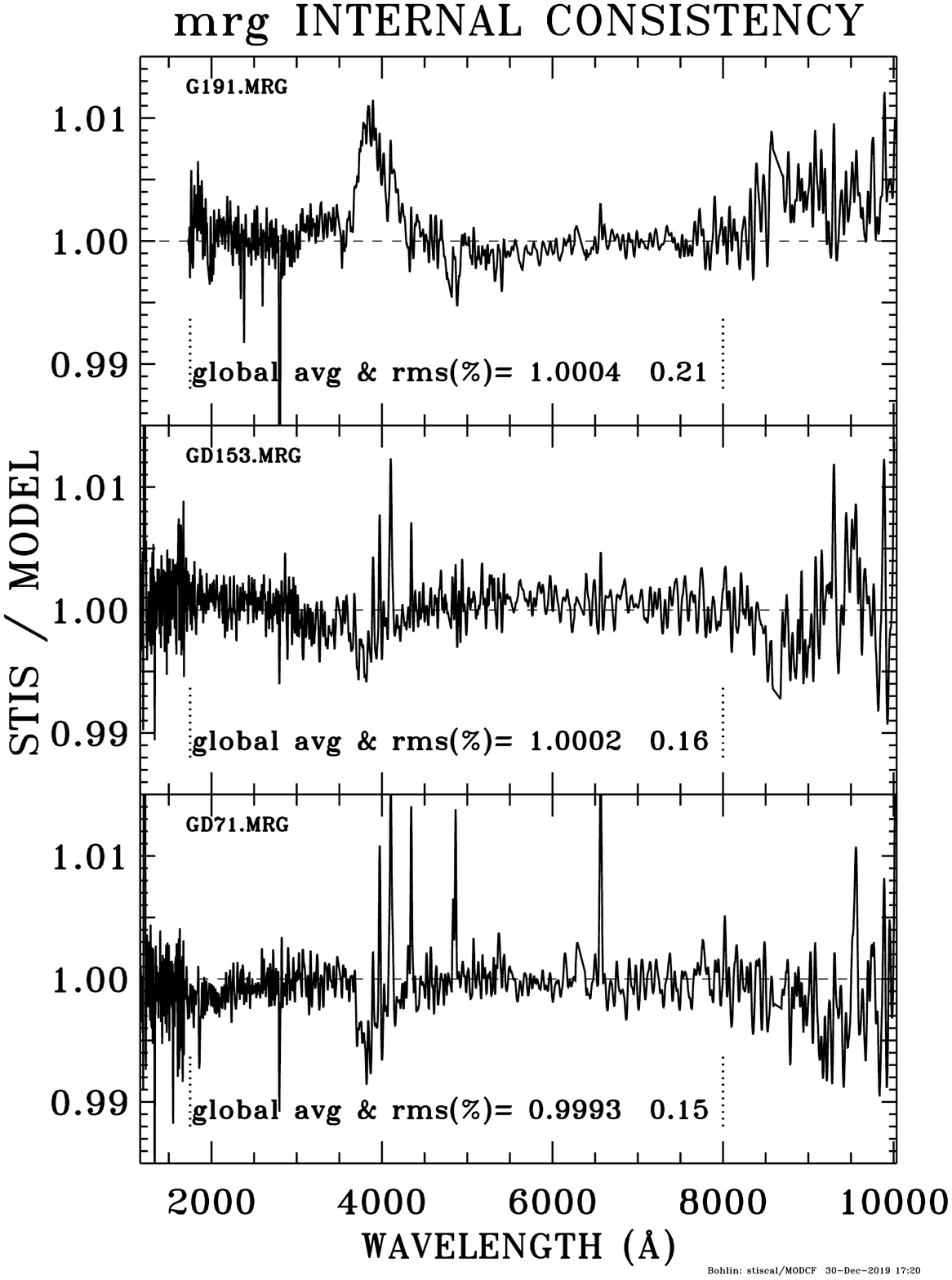}
\caption{\baselineskip=12pt
Residuals for our new set of primary standard WDs. STIS SEDs from the five
low-dispersion observing modes in the 52x2 slit are combined to make the
numerator flux distribution, while the denominators are the new model SEDs
calculations from \textsc{tmap} for GD153 and GD71 and from \textsc{tlusty} for
G191-B2B. The average and rms values written on each panel are computed in the
range delineated by vertical dotted lines at 1750~\AA\ and 8000~\AA.
\label{modcfmrg}}
\end{figure}

As a final confirmation of the consistency of our three prime standards,
Figure~\ref{modcfmrg} shows that the ratio of the calibrated data to the chosen
model flux has broadband residuals of $<$0.5\% except in the hard-to-model
confluence of the Balmer lines near 3800~\AA\ for G191-B2B. These residuals are
slightly improved in comparison to the previous pure \textsc{tmap} model set for
the three primary standard WDs that is shown in figure 13 of \citet{Bohlin2019}.
Using Sloan spectroscopy and photometry, \cite{aprieto2009} show a similar graph
in their figure 8, where their broadband residuals sometimes approach 1.5\% in
the 3000--7000~\AA\ range. \cite{aprieto2009} have a different model grid,
different assumptions about interstellar reddening, and a pure hydrogen model
for the metal line-blanketed G191-B2B. However, our new self-consistent analysis
demonstrates residuals generally $<$0.5\% in the same wavelength range, with the
exception of previously mentioned narrow bump near 3800~\AA\ for G191-B2B.

\subsection{Saturated CCD Data for Vega and Sirius}	

\begin{figure}			
\centering 
\includegraphics*[width=.99\textwidth,trim=0 65 0 55]{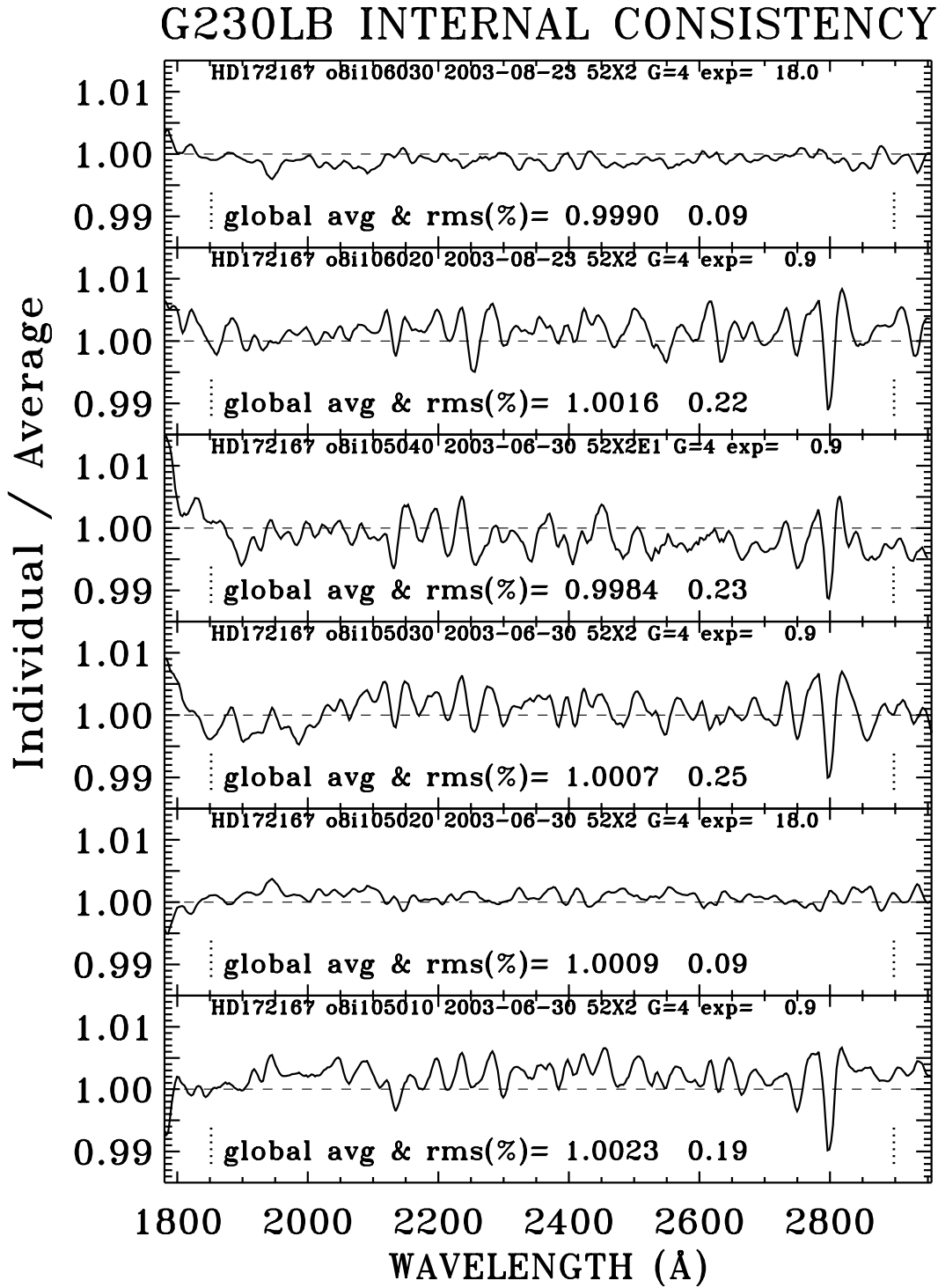}
\caption{\baselineskip=12pt
Residuals for the short (0.9~s) unsaturated and long (18~s) saturated
observations of Vega. The average and rms values are on the plots for the range
delineated by vertical dotted lines at 1750~\AA\ and 2950~\AA. The strongest
features in the short exposure ratios are 2800~\AA\ Mg II absorption lines that
have slightly more contamination by out-of-band light in the tall extraction
height of 84 pixels required to catch all the saturation. \label{g230lbvega}}
\end{figure}

Historically, two of the brightest stars in the sky, Sirius and Vega, have been
extensively observed for purposes of measuring their absolute flux
distributions. STIS can observe these important stars in the well-calibrated CCD
low-dispersion CCD modes G230LB, G430L, and G750L \citep{riley2019}.
Unfortunately, these standard stars are so bright that all the CCD data are
saturated, except for Vega with G230LB and the shortest reliable exposure time.
However, \citet{gillil2004} demonstrated that the response is still linear to
0.1\% for a factor of 50 overexposure, as long as all the saturated pixels are
included in the spectral extractions height. At the gain=4 setting, the excess
charge just bleeds into adjacent pixels along the columns, which are
perpendicular to the dispersion axis. As detailed by  \citet{bohlin2004} and
\citet{bohlin14}, the net signal extracted with heights larger than the standard
11 pixels includes an extra fraction of the PSF, which must be removed before
applying the standard flux calibration procedure. This extra signal is the ratio
of the large height to the 11 pixel net signal derived from unsaturated
observations of AGK+81$^{\circ}$266 in each of the CCD modes with correction
for CTE losses in both heights. As proof of this procedure the unsaturated
0.9~s G230LB SEDs of Vega are compared to the flux derived from the saturated
18~s exposures in Figure~\ref{g230lbvega}. The short exposure fluxes agree with
the long exposures to 0.23\%. The wide extractions of the saturated data have
such a large signal that the CTE correction is negligible; formally, the CTE
correction formula of \citet{goudfrooij2006} predicts $<$0.1\% loss.

A previously unaccounted complication in defining the correction for the large
extraction  heights is that the spectral widths are increasing over time
\citep{Bohlin2019,bohlin15}. Thus, the Vega corrections are the average of the
13 AGK+81$^{\circ}$266 observations within a year of the 2003.6 epoch of the
Vega data, while more coverage in the timeframe of the Sirius data at the
average 2012.9 date permits a window of $\pm$3 years. The uncertainty in these
corrections for the large heights is estimated from the difference between the
results for G230LB from the unsaturated Vega vs. AGK+81$^{\circ}$266, which 
range from 1.1\% at 1720~\AA\ to 0.6\% at 3050~\AA. Because
the signal is the largest at 3050~\AA\ and closer to the typical signal levels
of G430L and G750L, perhaps, the 0.6\% uncertainty is appropriate for the longer
wavelengths.

\newpage

\begin{figure}			
\centering 
\includegraphics*[width=.99\textwidth,trim=0 30 0 55]{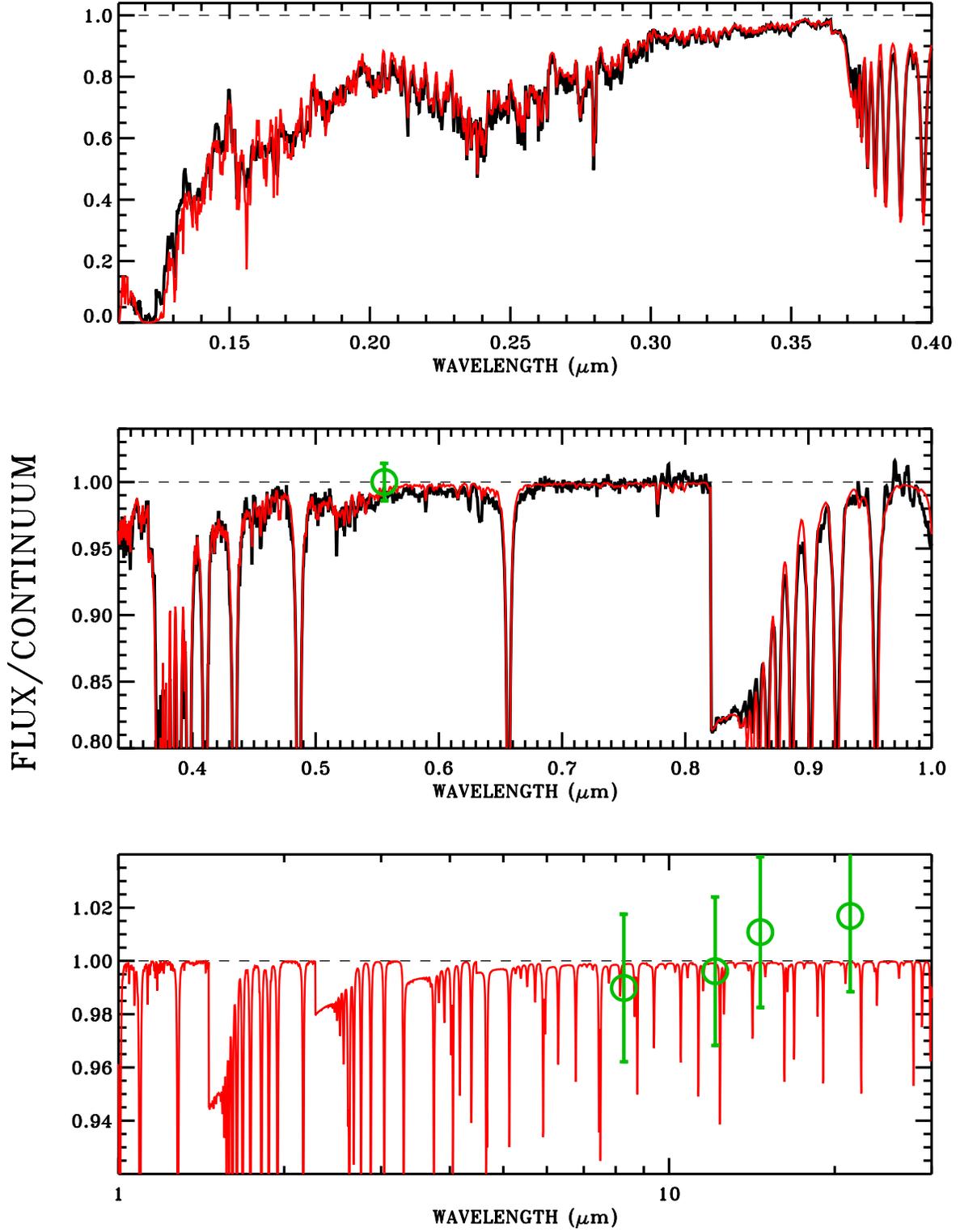}
\caption{\baselineskip=12pt Comparison of STIS (black) to the Sirius model. The
model and its continuum are both normalized to the STIS SED in the
6800--7700~\AA\ region. The green circles are the MSX values of \citep{price04}
and the \citet{megessier95} $3.46\times10^{-9}$~erg cm$^{-2}$ s$^{-1}$ \AA$^{-1}$
at unity that determined the original normalization of the CALSPEC system. 
Notice the expanded scale in the bottom panel. \label{sirius}} \end{figure}

Figure~\ref{sirius} compares the STIS flux for Sirius to a R = 500 resolution
Kurucz model. Below 1675~\AA, an International Ultraviolet Explorer (IUE)
spectrum completes the measured SED after normalization to the STIS flux by
multiplying by 1.26. In order to display the spectral features on an expanded
scale, the illustrated SEDs have been divided by the same theoretical smooth
continuum. The model and the continuum are both normalized to the STIS flux in
the line-free 6800--7700~\AA\ continuum region, where the model should be the
most reliable. Ratio plots of STIS/model have distracting, spurious dips and
spikes near absorption line features because of small mismatches in resolution,
tiny wavelength errors, and uncertain line strengths. The ratios of
flux/continuum in Figure~\ref{sirius} display the nature of the absorption
lines, and the irrelevant small mismatches at line centers are often off-scale
and can be easily ignored. Many of the weaker STIS absorption lines have
corresponding features in the Kurucz model, so the model replicates the
measurements with high fidelity.  Thus, longward of 1~\micron, the 2013 Kurucz
model (private communication), normalized to the STIS flux in the
6800--7700~\AA\ range, is chosen for the composite CALSPEC SED for Sirius.  In
contrast to Vega with its unresolved dust ring, Sirius has no dust ring and this
composite SED of STIS plus model  should comprise a good fundamental standard
reference spectrum to at least 30~\micron. On the other hand, the  Vega dust
ring dominates the far-IR SED and prohibits the use of a photospheric model as
an IR flux standard. The IR emission from the Vega dust contributes more than
1\% to the photospheric flux longward of $\approx$2~\micron, e.g.,
\citet{bohlin14}, so the CALSPEC photospheric model for Vega in the IR does not
represent the true SED for unresolved observations of the total, photosphere
plus dust disk.

Following the logic of \citet{bohlin14}, the four green circles in the IR in
Figure~\ref{sirius} are the ratio of the MSX results of \citep{price04} to the
Sirius model, which is normalized to the STIS SED that uses the original
\citet{megessier95} Vega flux of $3.46\times10^{-9}$~erg cm$^{-2}$ s$^{-1}$
\AA$^{-1}$ at  5557.5~\AA\ (vac) to set the absolute CALSPEC flux scale. The
average MSX ratio is 1.0034, in contrast to the previous value of 0.989
\citep{bohlin14}. With a unit ratio at 5557.5~\AA\ for the original
CALSPEC absolute flux zero point, the average of the visible and IR absolute
flux ratios is 1.0017 too high; thus, the CALSPEC fluxes must all be increased
by this factor to reconcile the two sets of preferred absolute flux measures
that have the same weight. Equivalently, the CALSPEC absolute flux levels are
set by increasing the \citet{megessier95} value to $3.47\times10^{-9}$~erg
cm$^{-2}$ s$^{-1}$ \AA$^{-1}$ at 5557.5~\AA\ (vac, 5556~\AA\ air), which
happens to agree
exactly with \citet{tug1977}. The formal uncertainty is $\pm$0.5\% in the
average from combining the Megessier uncertainty of 0.7\% with the total
uncertainty of 0.7\% for the set of four MSX points. This new absolute level for
Vega replaces the previous $3.44\times10^{-9}$~erg cm$^{-2}$ s$^{-1}$ \AA$^{-1}$
of \citet{bohlin14}. This gray increase of 0.87\% (3.47/3.44) in all the CALSPEC
fluxes is in addition to the wavelength-dependent changes in
Figure~\ref{flxchang} that are due to improved WD model SEDs.

\subsection{CALSPEC vs. \citet{tug1977} Absolute Flux Measures}		

\begin{figure}			
\centering 
\includegraphics*[width=.99\textwidth,trim=0 25 0 15]{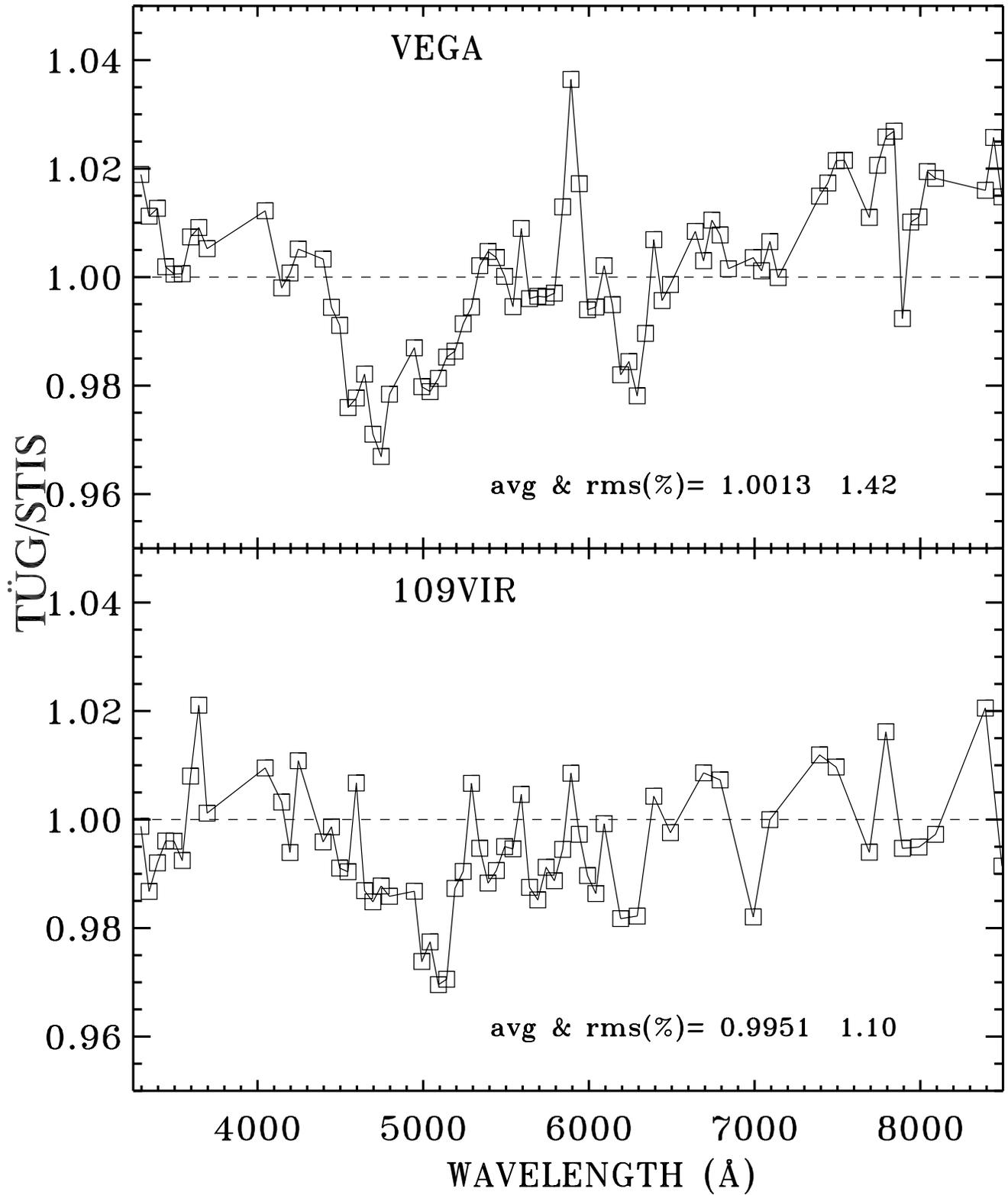}
\caption{\baselineskip=12pt
Comparison of the CALSPEC SEDs with the independent absolute flux measures of 
\citet{tug1977} for Vega and 109 Vir. The CALSPEC flux distributions are validated to 1--2\% in broad bands.
\label{tug}}
\end{figure}

Now that 109 Vir has been observed by STIS, the CALSPEC fluxes can be compared
to the absolute flux determinations for both Vega and 109 Vir in
\citet{tug1977}. Figure~\ref{tug} shows the ratio of the \citet{tug1977} results
to the CALSPEC \textit{alpha\_lyr\_stis\_010.fits} and
\textit{109vir\_stis\_002.fits} SEDs. \citet{tug1977} cover the 3295--9040~\AA\
region with 10~\AA\ bandpasses at the shorter wavelengths and 20~\AA\ bandpasses
at the longer wavelengths. In general, \citet{tug1977} measure at continuum
wavelength sample points that avoid the confusion of absorption lines. However,
that policy was enforced only at wavelengths below 8500~\AA, which invalidates
the ratios in the Paschen line region where resolution differences and tiny
wavelength errors cause excess noise. Thus, Figure~\ref{tug} shows only the
valid 3295--8500~\AA\ region with the average ratio and rms written on the
plots. The ratios and rms are roughly consistent with the \citet{tug1977}
uncertainty of 1\% longward of 4000~\AA, where most of the points are within
2$\sigma$ of unity, and only the one point for Vega at  5890~\AA\ deviates by as
much as 4$\sigma$. The two independent sets of absolute fluxes agree better than
the comparison of CALSPEC with \citet{hayes1985} in the visible or with
comparisons in the ultraviolet that are all shown in \citet{bohlinetal14}. None
of these comparisons show evidence of systematic CALSPEC flux errors. 

\newpage

\section{Fitting Pure Hydrogen Models to STIS Observations}	

Following the \citet{bohlinetal14} and \citet{bohlinsed2019} reduced $\chi^2$
method of fitting K to O type stars, WD SEDs are fit using the new
\textsc{tlusty207} and \textsc{tmap2019} pure hydrogen grids discussed in
Section 2. The extrapolated SEDs from the 1~\micron\ STIS long-wavelength cutoff
to 32~\micron\ serve as reference flux distributions for IR flux calibration
with a firm HST anchor at the shorter wavelengths. Any HST IR grism
spectrophotometry that reaches to the Wide Field Camera 3 (WFC3) 1.7~\micron\
limit or to the Near Infrared Camera and Multi-Object Spectrograph (NICMOS)
2.5~\micron\ limit increases the lever  arm of the extrapolation and decreases
uncertainties. Both the WFC3 and NICMOS grism flux calibrations have been
revised using the new SEDs for the three primary standards. The parameters of
the best reduced $\chi^2$ fits are $T_\mathrm{eff}$, $\log g$, and E(B-V) for
each modeled WD, as listed in Table~\ref{table:wds}. The maximum residual
difference between the model fits and the data for the hotter stars does not
exceed 1.6\% in broad bins, except for SDSSJ151421, where the maximum residual
is 2.5\% in the 3300--3700~\AA\ bin. The CALSPEC WD SEDs concatenate the
measurements with the model fits at the longer wavelengths.

\begin{figure}			
\centering 
\includegraphics*[width=.99\textwidth,trim=0 30 0 15]{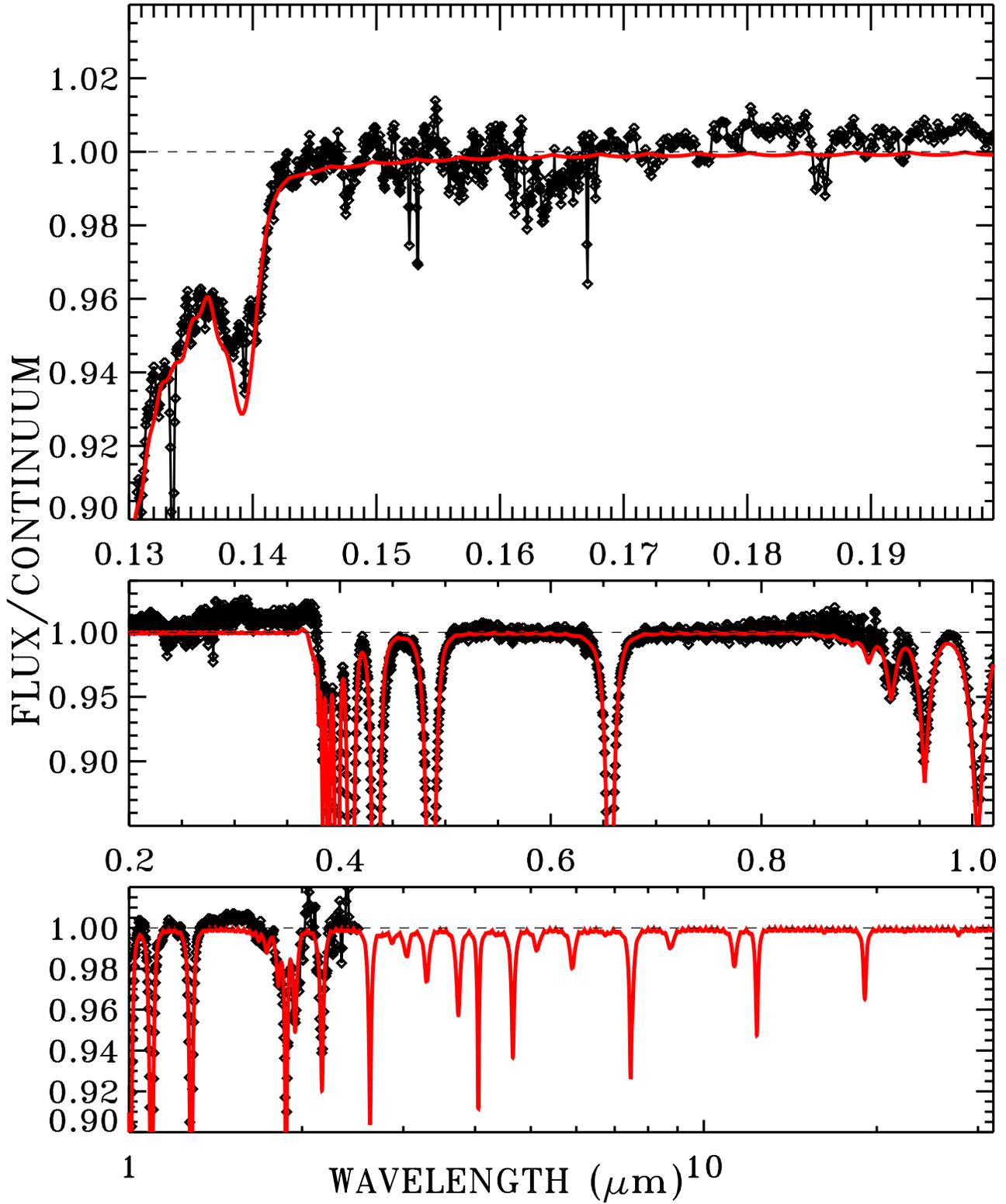}
\caption{\baselineskip=12pt
Comparison of the measured SED for GRW+70$^{\circ}$5824 to our \textsc{tlusty} model (red). The observed SEDs and the model are divided by the
model continuum for the Table~\ref{table:wds} fit. \label{grw}} \end{figure}

As an example of the utility of models that include quasi-molecular satellites
(QMS), Figure~\ref{grw} compares the strong observed QSM feature at 1400~\AA\ to
the \textsc{tlusty} model for the cool WD GRW+70$^{\circ}$5824. Despite
confusion by absorption from the Si IV ground state at 1393.8 and 1402.8~\AA,
the \textsc{tlusty} model illustrates the nature of the QMS. While there are
metal lines that could invalidate the fit, the model approximates the observed
SED within $\approx$2\% longward of 1300~\AA. However, the metal-line features,
the broad hydrogen lines, and the small systematic deviations of the model from
the CALSPEC SED \textit{grw\_70d5824\_stiswfcnic\_002.fits} suggest that the
model for GRW+70$^{\circ}$5824 does not make an ideal standard star. However,
the observed SED \textit{grw\_70d5824\_stiswfcnic\_002.fits} is one of the best
CALSPEC standards.

\begin{deluxetable}{lcccc}		
\tablewidth{0pt}
\tablecaption{\label{table:wds} Parameters of the Model Fits}
\tablehead{
\colhead{Star} &\colhead{$T_\mathrm{eff}$} &\colhead{$\log g$}
&\colhead{E(B-V)} &$\chi^2$}
\startdata
GRW+70$^{\circ}$5824\tablenotemark{a} & 20540  &7.90 & 0.000 &0.05  \\
SDSSJ151421                          & 29120  &8.90 & 0.043 &0.64  \\
WD0320-539                           & 33110  &7.60 & 0.000 &0.10  \\
WD0947+857                           & 40020  &7.55 & 0.000 &0.25  \\
WD1026+453                           & 35240  &7.55 & 0.000 &0.20  \\
WD1657+343\tablenotemark{a}          & 46750  &7.10 & 0.000 &0.10  \\
\enddata
\tablenotetext{a}{WFC3 and NICMOS grism data extend the observed SED longward
of 1~\micron.}
\tablecomments{Results from fitting model atmospheres to the observed stellar
SEDs using $\chi^2$ fitting with the \textsc{tmap} pure hydrogen grid, except
for GRW+70$^{\circ}$5824, which is fit with the \textsc{tlusty} grid. The
parameters of the fit for each star are the effective temperature
$T_\mathrm{eff}$, the surface gravity $\log g$, the interstellar reddening
E(B-V), and the reduced chi-square quality of the fit $\chi^2$.} \end{deluxetable}

\subsection{CALSPEC vs. \citet{narayan2019}}			

\begin{figure}			
\centering 
\includegraphics*[width=.99\textwidth,trim=0 20 0 15]{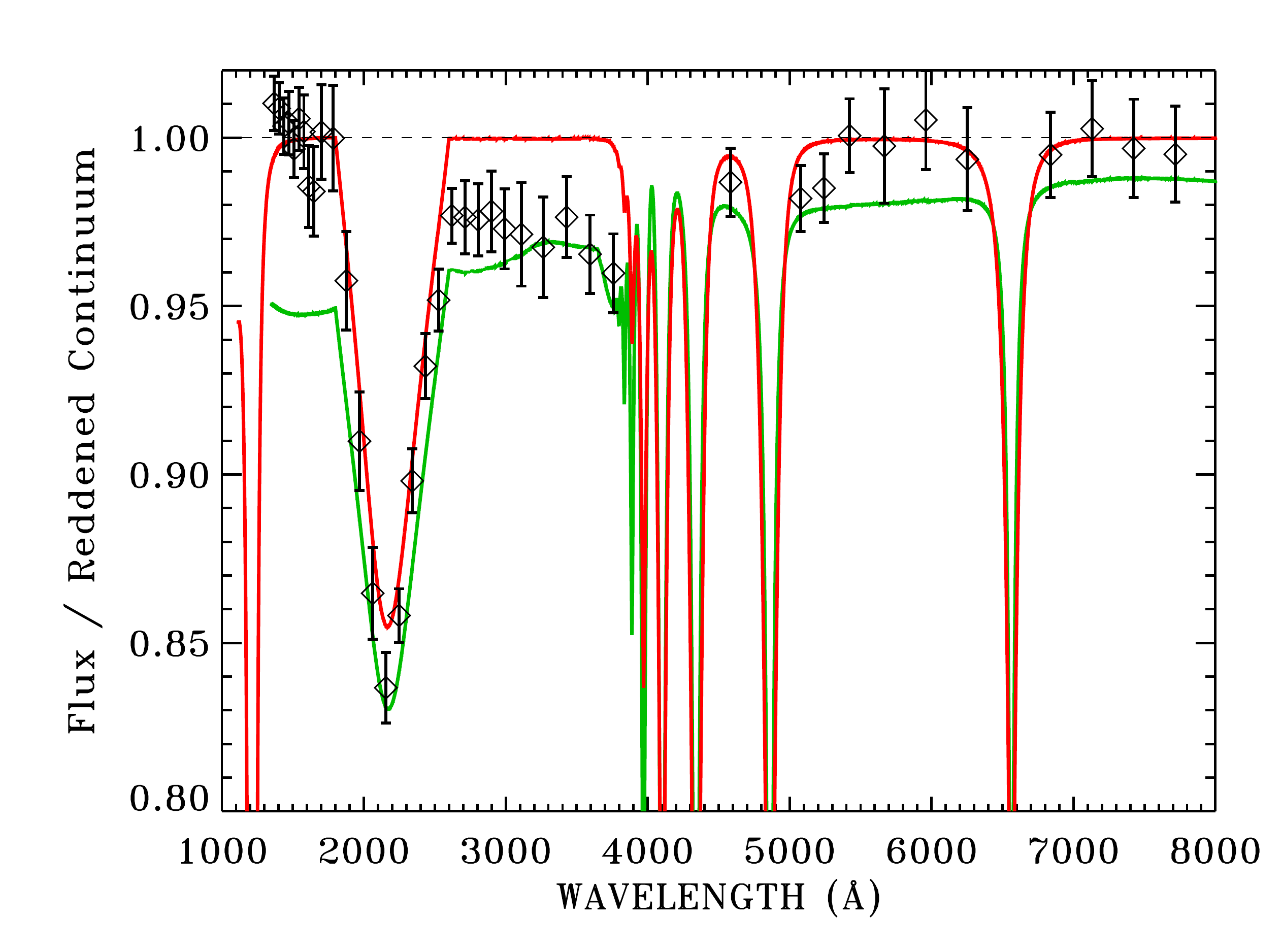}
\caption{\baselineskip=12pt
Comparison of the binned STIS CALSPEC SED for SDSSJ151421.27+004752.8
(black diamonds and 2$\sigma$ error bars) with the pure hydrogen model fit to
WFC3 photometry by
\citet{narayan2019} (green) and to our model (red) that best fits the STIS
observation. The black STIS SED uses the same old basis for the flux calibration
as \citet{narayan2019}. The observed SEDs and both models are divided by the
model continuum for the STIS fit with the 2200~\AA\ bump removed by linearly
interpolating the continuum from 1800 to 2600~\AA. The red model deviates from
the STIS SED in the depth of the 2200~\AA\ feature and by $\approx$3\% in the
2600--3800~\AA\ region, which is indicative of a reddening curve that differs from the average. \label{stiscf}} \end{figure}

\citet{narayan2019} and \citet{calamida2019} have produced an all-sky network of
faint WD absolute flux standards that are tied to the CALSPEC absolute flux
scale by fitting F275W, F336W, F475W, F625W, F775W, and F160W filter photometry
from the WFC3 on HST. One star SDSSJ151421.27+004752.8 in the network is bright
enough at V = 16.5 to obtain fair S/N with STIS in two orbits and was observed
24 April 2019. The best fit to the STIS SED from Table~\ref{table:wds} is a
model with $T_\mathrm{eff}$=29120~K, $\log g$=8.90, and E(B-V)=0.043 for a
calibration based on the new set of three primary flux standards. However, to
make a fair comparison with \citet{narayan2019}, who based results on the
previous system, a fit of the old SED for SDSSJ151421
\textit{grw\_70d5824\_stiswfcnic\_001.fits} is used for Figure~\ref{stiscf} with
its fit at $T_\mathrm{eff}$=29010~K, $\log g$=8.85, and E(B-V)=0.043.
Figure~\ref{stiscf} compares this old SED to the \citet{narayan2019} fit of
$T_\mathrm{eff}$=28768~K, $\log g$=7.89, and E(B-V)=0.039. Differences in these
model labels are inconsequential; the important difference is between the pure
hydrogen model flux distributions that are proposed as standard reference SEDs,
as illustrated in Figure~\ref{stiscf}. The old STIS SED (black diamonds with
2$\sigma$ error bars), its \textsc{tmap2019} model (red), and the
\citet{narayan2019} model (green) are all divided by the \textsc{tmap2019}
reddened continuum. In order to show the main spectral features, including the
2200~\AA\ extinction bump, the reddened continuum is interpolated linearly
across the 1800--2600~\AA\ range, revealing a $\approx$15\% dip at 2200 in the
red model fit. The STIS observation is binned into statistically independent
bins of 60 pixels and the error-in-the-mean is computed from the rms scatter
among the 60 samples in each of bins. These error bars computed from the data
are typically 2--4 times larger than the propagated statistical errors from the
Poisson statistics.

The data are  convincingly lower than the
red model fit by 2--3\% at 2200--3800~\AA, where the measures agree well with
the green fit but suggest that the reddening is anomalously larger than the
average reddening used for the red model.  These small deviations of the STIS
SED from its red model are caused in part, or maybe in full, by typical
variations from the \citet{cardelli89} average reddening assumed in the fitting
process. Using other average reddening curves such as  \citet{fitzpatrick1999}
does not significantly change the results. For illustrations of the range of
individual interstellar extinction curves, see \citet{witt1984}.

At wavelength longer than 5000~\AA, the systematically larger STIS flux in all
ten independent bins suggests that the green \citet{narayan2019} fit is 1--2\%
low, which is a bit surprising, as the \citet{narayan2019} fit is well
constrained by WFC3 photometry in that wavelength region. Possibilities for the
general trend of the green curve to fall below the red are errors in the STIS
spectrophotometry of the faint SDSSJ151421.27+004752.8 and problems in
accurately placing the WFC3 photometry on the CALSPEC scale, including small
errors in the lab measurements of the filter throughput curves.
\citet{bohlin2016} found that some ACS filter curves required significant
changes in order to achieve a 1\% photometry goal. The STIS observations of
SDSSJ151421 should be repeated to confirm the single measurement for such a
faint star.

In summary, the \citet{narayan2019} results are confirmed in the region where
there are WFC3 constraints at the F275W filter and longward, but only to a
precision of 2\%, rather than to the goal of 1\%. At FUV wavelengths in the
unconstrained 1500~\AA\ region, the fit of \citet{narayan2019} is too low
by $\approx$5\%.

Other differences between the red and green curves in Figure~\ref{stiscf}
include the use of an earlier version of the \textsc{tlusty} grid, rather than
the new \textsc{tmap2019} grid recommended here. For the best absolute fluxes of
the all-sky network of faint WDs, a new analysis should include a recalibration
of the WFC3 photometry using the three revised primary WD flux standards,
followed by a re-fitting of that recalibrated WFC3 photometry with the new
\textsc{tmap2019} grid. The uncertainties associated with anomalous reddening
curves in the UV and small filter bandpass shifts should also be quantified.

\newpage
\section{Summary}	

Because of the changes in the SEDs of the three primary standards, all of the
HST flux calibrations must be updated to achieve the goal of $\approx$1\%
precision.. The recalibration of the five STIS low dispersion modes and the WFC3
and NICMOS IR grisms is complete, and the 99 stars in the CALSPEC database with
STIS, WFC3, or NICMOS observations are re-delivered. In addition, models for 73
of the 99 stars are included in the update. No models are available for the
three stars with observations of NICMOS only, for five M, T, or L type stars, or
for the stars with partial STIS coverage in table 1b of CALSPEC. The
remaining 16 stars with no model include cases where there is a dust ring that
affects the SED in the IR, where the WD is too cool or is not pure hydrogen,
where the star is variable, or where there is a close companion that affects the
measured SED. The model files for the main-sequence stars are the R=300,000 BOSZ
models of \citet{bohlin2017}, while the R=500 BOSZ version is concatenated to
the observed SED at the long wavelengths of the CALSPEC data files. The
concatenated portion of the WD data files is the same as in the model WD files.

The changes in flux for the updated SEDs often do not correspond exactly to the
average change in Figure~\ref{flxchang} plus the 0.87\% gray increase of Section
3.1, because of the non-uniform processing dates of the replaced files. Other
smaller changes occured over time, e.g., increasing the extraction width from 7 to
11 pixels for the CCD data and updating the correction for sensitivity changes
over time, etc. \citep{Bohlin2019}.

The new WD pure hydrogen grids are available in the MAST Archive~via
\dataset[https://doi.org/10.17909/t9-7myn-4y46]{https://doi.org/10.17909/t9-7myn-4y46}
.


\newpage

\section*{Acknowledgements}

Support for this work was provided by NASA through the Space Telescope Science
Institute, which is operated by AURA, Inc., under NASA contract NAS5-26555.
Thanks to Abi Saha for inspiring this paper. His team has established a network
of faint WD spectrophotometric flux standards, and the team members are
the author lists of \citet{narayan2019} and \citet{calamida2019}.

~\\
\textbf{ORCID IDs}

Ralph C. Bohlin https://orcid.org/0000-0001-9806-0551

Ivan Hubeny https://orcid.org/0000-0001-8816-236X

Thomas Rauch https://orcid.org/0000-0003-1081-0720

\bibliographystyle{apj}

\bibliography{../../../pub/paper-bibliog}

\end{document}